\documentclass[aps, pre, twocolumn, 10pt, amsmath, amssymb, amsfonts]{revtex4-1}

\usepackage                   {graphicx}
\usepackage[usenames]         {color}
\usepackage[colorlinks=True,
            urlcolor=myblue,
            linkcolor=myblue,
            citecolor=mygreen]{hyperref}
\definecolor{mygreen}{RGB}{ 0, 127,   0}
\definecolor{myblue} {RGB}{ 0,   0, 127}
\newlength{\figwidth}
\newcommand{\fig}[1]{Fig.~\ref{#1}}
\newcommand{\eq}[1]{Eq.~(\ref{#1})}
\newcommand{\ud}{\,\mathrm{d}}
\newcommand{\vect}[1]{\mathbf{#1}}

\begin{document}
\title{Invariants in the Yukawa system's thermodynamic phase diagram}
\author{Arno A. Veldhorst}
\email{a.a.veldhorst@gmail.com}
\altaffiliation{Current Address: Laborat\'orio de Espectroscopia Molecular, Instituto de Qu\'imica, Universidade de S\~ao Paulo, CP 26077, CEP 05513-970 S\~ao Paulo, SP, Brazil}
\affiliation{DNRF Centre ``Glass and Time'', IMFUFA, Dept. of Sciences, Roskilde University, P.O. Box 260, DK-4000 Roskilde, Denmark}
\author{Thomas B. Schr{\o}der}
\affiliation{DNRF Centre ``Glass and Time'', IMFUFA, Dept. of Sciences, Roskilde University, P.O. Box 260, DK-4000 Roskilde, Denmark}
\author{Jeppe C. Dyre}
\email{dyre@ruc.dk}
\affiliation{DNRF Centre ``Glass and Time'', IMFUFA, Dept. of Sciences, Roskilde University, P.O. Box 260, DK-4000 Roskilde, Denmark}
\date{\today}

\begin{abstract}
This paper shows that several known properties of the Yukawa system can be derived from the isomorph theory, which applies to any system that has strong correlations between its virial and potential-energy equilibrium fluctuations. Such ``Roskilde-simple'' systems have a simplified thermodynamic phase diagram deriving from the fact that they have curves (isomorphs) along which structure and dynamics in reduced units are invariant to a good approximation. We show that the Yukawa system has strong virial potential-energy correlations and identify its isomorphs by two different methods. One method, the so-called direct isomorph check, identifies isomorphs numerically from jumps of relatively small density changes (here 10\%). The second method identifies isomorphs analytically from the pair potential. The curves obtained by the two methods are close to each other; these curves are confirmed to be isomorphs by demonstrating the invariance of the radial distribution function, the static structure factor, the mean-square displacement as a function of time, and the incoherent intermediate scattering function. Since the melting line is predicted to be an isomorph, the theory provides a derivation of a known approximate analytical expression for this line in the temperature-density phase diagram. The paper's results give the first demonstration that the isomorph theory can be applied to systems like dense colloidal suspensions and strongly coupled dusty plasmas.
\end{abstract}
            
\maketitle

\section{Introduction}
The Yukawa pair potential has been used to model a wide variety of different phenomena in physics~\cite{Rowlinson1989}. Named after Hideki Yukawa who used this potential in his meson theory~\cite{Yukawa1935}, Debye and H{\"u}ckel~\cite{Debye1923} had in fact used it earlier to describe the interactions between ions in solutions. The Yukawa pair potential, which is also referred to as the screened Coulomb potential, has the form

\begin{equation}\label{eq:y_debyuehuckel}
  \upsilon(r) = \dfrac{Q^2}{r} \,\, \exp\left(-r/\lambda\right)\,.
\end{equation}
Here $Q$ is the particle charge in the Gaussian unit system and $\lambda$ the so-called screening length. Although Debye-H{\"u}ckel theory was devised for describing the behavior of dilute ionic solutions, the Yukawa potential has proven useful also in the description of charge carriers that are much larger than those of the surrounding medium. This is the case for instance in suspensions of charge-stabilized colloids or in dusty plasmas, compare, e.g. the DLVO theory~\cite{Derjaguin1941, VerweyOverbeek} in which the interaction between the surfaces of two colloids is described by a potential of the Yukawa form. The colloid particles are often modeled with a potential that adds a hard-core repulsion, but the low-temperature part of the phase diagram of this ``hard-core'' Yukawa potential can be mapped onto the phase diagram of the Yukawa pair potential~\cite{Hynninen2003}.

Besides electrons and ions, dusty plasmas contain small solid particles that are charged. In dusty plasmas the distance between the solid particles is usually large compared to the size of the particles, and their interaction can therefore be modeled using the (point) Yukawa potential of Eq. (\ref{eq:y_debyuehuckel}). A more involved potential consisting of a sum of Yukawa terms is sometimes used to capture phenomena like plasma production and loss balance~\cite{Khrapak2010}. The physics of dusty plasmas is not only of interest for industrial applications where such plasmas are formed, but also in astrophysics for the understanding of stellar materials and planet formation~\cite{Fortov2005, Morfill2009}.

In 1986 it was found that colloids in suspension can form crystal lattices~\cite{Pusey1986} and it was predicted that this should also be possible in dusty plasmas~\cite{Ikezi1986}. This created a renewed interest in the phase diagram of the Yukawa system~\cite{Robbins1988, Meijer1991, Farouki1992}. It was later found in experiments that plasma crystals indeed do exist~\cite{chu1994, Thomas1994}. If $\rho$ is the number density of particles, in the field of dusty plasmas the phase diagram is usually presented in terms of the following two dimensionless parameters ~\cite{Dzhumagulova2013, Yazdi2014}: the screening parameter
\begin{equation}\label{scr}
\kappa\equiv\frac{\rho^{-1/3}}{\lambda}
\end{equation}
and the coupling parameter
\begin{equation}\label{coup}
\Gamma\equiv \frac{Q^2\,\rho^{1/3}}{k_BT}\,.
\end{equation}
Physically, the screening parameter is much larger than unity whenever the screening length is much smaller than the average interparticle distance -- this characterizes the part of the thermodynamic phase diagram where the exponential damping term dominates over the Coulomb term. In the other limit, $\kappa\ll 1$, the exponential damping term plays little role and the system behaves as a single-component repulsive Coulomb system, the so-called one-component plasma (OCP), with an ill-defined thermodynamic limit due to the infinite screening length. The case $\Gamma\gg 1$ corresponds to the potential energy from the individual pair interactions being much larger than the thermal energy; this favors crystallization, depending on the value of $\kappa$, see below. In the opposite limit, $\Gamma\ll 1$, the system approaches an ideal gas. Care must be taken when comparing $\kappa$ or $\Gamma$ values in the literature, because they often include in their definition further multiplicative constants. For instance, in the above definitions of the screening and coupling parameters the term $\rho^{-1/3}$ is often replaced by the Wigner-Seitz radius $a=[3/(4\pi\rho)]^{1/3}$. 

The Yukawa system's phase diagram is well understood with its two solid phases and a fluid phase~\cite{Ikezi1986, Robbins1988, Meijer1991, Farouki1992}. At high $\kappa$ the Yukawa fluid crystallizes into an face-centered cubic (FCC) lattice, while at low $\kappa$ it crystallizes into a body-centered cubic (BCC) lattice. The triple point separating fluid, FCC, and BCC phases is located at $\kappa\cong 6.90, \Gamma \cong 3.47\times 10^3$~\cite{Hamaguchi1997}. Vaulina et al.~\cite{Vaulina2000, Vaulina2002} derived an expression for the melting line from the pair potential using the Lindemann melting criterion \cite{Lindemann1910, Gilvarry1956}. This expression is confirmed below, where it is derived from the isomorph theory. This theory applies to the class of systems (dense liquids and classical crystals) termed \emph{Roskilde-simple} -- or just \emph{Roskilde} (\emph{R}) -- systems~\cite{Ingebrigtsen2012c, Dyre2014}. As documented below, the Yukawa system exhibits the strong correlations between equilibrium virial and potential-energy fluctuations required for a system to belong this class. 

Over the years, investigations of the dynamics of the Yukawa system have mostly focused on the self-diffusion coefficient and the viscosity~\cite{Robbins1988, Kremer1987, Lowen1993, Rosenfeld2000, Rosenfeld2001, Ohta2000, Daligault2012, Khrapak2012, Dzhumagulova2013, Faussurier2014}. In the present publication we derive some of the previously reported scaling properties of the Yukawa system's transport coefficients from the general isomorph theory.

\begin{figure}
  \centering
  \includegraphics[width=\figwidth]{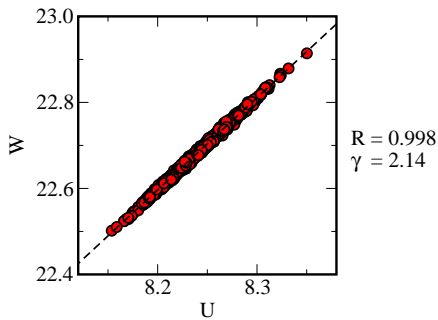}
  \caption{Scatter plot (red symbols) demonstrating strong correlations between the instantaneous values of the virial $W$ and the potential energy $U$ during an equilibrium $NVT$ simulation of the Yukawa system at $\rho = 3\times 10^{-3}$ and $T=5.3608\times 10^{-5}$ ($\kappa=6.934$ and $\Gamma=2690$). The dashed line is the standard linear regression with slope $\gamma$.}\label{fig:UWscatter}
\end{figure}

The isomorph theory initiated from the observation that some model liquids have strongly correlated fluctuations of their energy and pressure~\cite{Pedersen2008b}. More precisely, the correlations are between the {\it configurational} parts of these quantities. If one for a system of $N$ particles splits energy and pressure into ideal-gas terms, which depend only on the particle momenta, and configurational terms depending only on the particle positions as follows
\begin{align}
  E  &= K(\vect{p}_1, \ldots, \vect{p}_N) + U(\vect{r}_1, \ldots, \vect{r}_N)\,,\label{eq4}\\
  pV &= N k_B T(\vect{p}_1, \ldots, \vect{p}_N) + W(\vect{r}_1, \ldots, \vect{r}_N)\,,\label{eq5}
\end{align}
strong correlations between the potential energy $U$ and virial $W$ constant-volume thermal-equilibrium fluctuations are found for a number of systems~\cite{Pedersen2008}. An illustration of such strong virial potential-energy correlations is shown in \fig{fig:UWscatter} that plots the instantaneous values of $W$ versus those of $U$ during an $NVT$ equilibrium simulation of the Yukawa system. The correlations are quantified by the standard Pearson correlation coefficient~\cite{paper1}
\begin{equation}\label{eq:R}
  R = \frac{\left\langle \Delta W \Delta U \right\rangle}
           {\sqrt{\left\langle(\Delta W)^2\right\rangle \left\langle(\Delta U)^2\right\rangle}}\,.
\end{equation}
As \fig{fig:UWscatter} shows, $R$ is high for the Yukawa system. A system is generally considered to have strong $U,W$ correlation whenever $R>0.9$~\cite{paper1}, a convenient but also somewhat arbitrary criterion. Many model systems have been shown to exhibit strong $U,W$ correlations~\cite{paper1, Veldhorst2012, Bacher2014}, including the Lennard-Jones (LJ) system and other simple liquids, molecular liquid models~\cite{Ingebrigtsen2012b, Veldhorst2014}, liquids under shear~\cite{Separdar2013}, and crystals~\cite{Albrechtsen2014}. These systems were initially called ``strongly correlating'', but that repeatedly led to confusion with strongly correlated quantum systems, and they are now instead referred to as Roskilde-simple or just Roskilde (R) systems~\cite{Malins2013, Abramson2014, Henao2014, Pieprzyk2014, Prasad2014, Schmelzer2015, Buchenau2015, Heyes2015}. The isomorph theory has also been applied to nano-confined liquids~\cite{Ingebrigtsen2013b, Ingebrigtsen2014} and, e.g., to derive density-scaling invariants for zero-temperature plastic flow properties of glasses~\cite{Lerner2014}.

\begin{figure}
  \centering
  \includegraphics[width=\figwidth]{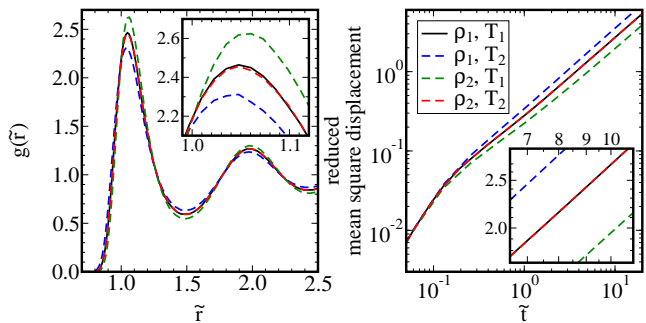}
  \caption{The radial distribution function (left) and the mean square displacement (right) in reduced units for four different state points, two of which are isomorphic to each other. The initial state point (black line) has $\rho_1=3\times 10^{-3}, T_1=5.361\times 10^{-5}$, while the isomorphic state point (dashed, red line) has $\rho_2=3.3\times 10^{-3}, T_2=6.553\times 10^{-5}$. The method used to identify the state point ($\rho_2,T_2$) as being  isomorphic to state point ($\rho_1,T_1$) is given below in Sec.~\ref{ssec:dic}. 
The two isomorphic state points are seen to have identical structure and dynamics to a very good approximation. The effects of either changing density or temperature (green and blue dashed lines) are shown for comparison. The insets show enlarged views of the same data.}\label{fig:smallstep}
\end{figure}

In 2009 it was found that R liquids have ``isomorphic'' curves in their thermodynamic phase diagram~\cite{paper4}. Isomorphs are curves along which all properties derived from structure or dynamics are invariant in properly reduced units, making the phase diagram effectively one dimensional with respect to many properties~\cite{paper4}. An example of this is shown in \fig{fig:smallstep}, where the Yukawa fluid is shown to have -- to a very good approximation -- identical radial distribution functions ($g(r)$) and mean square displacements at two state points that are isomorphic to each other. 

Having its origins in the field of organic, glass-forming liquids, the isomorph theory has been applied to explain certain empirical scalings in that field. These include the so-called density scaling, according to which the dynamics is a function of $\rho^\gamma/T$~\cite{Roland2005}. This is an approximation to the isomorph theory that applies when fairly small changes in density are considered, as in most experiments ~\cite{paper4, Bohling2012}. Because both the excess entropy, $S_{\rm ex}\equiv S-S_{ideal}$ ($S_{ideal}$ being the entropy of the ideal gas at same temperature and density), and the reduced-unit dynamics are invariant along isomorphs, the isomorph theory also predicts that R liquids obey Rosenfeld's excess entropy scaling, which states that the reduced transport coefficients are functions of $S_{ex}$~\cite{Rosenfeld1977}. Not predicted by the isomorph theory, but nevertheless found to apply to R liquids~\cite{Ingebrigtsen2013}, is the Rosenfeld-Tarazona relation~\cite{Rosenfeld1998} that gives expressions for the constant-density temperature dependence of the isochoric specific heat ($c_V\propto T^{-2/5}$). For a more comprehensive overview of the isomorph theory the reader is referred to a recently published Feature article~\cite{Dyre2014}.

The fact that the Yukawa system has strong $UW$ correlations shows that it belongs to the class of R liquids. This is consistent with the long-known facts that the Yukawa system~\cite{Rosenfeld2000, Sanbonmatsu2001} and the OCP~\cite{Daligault2006} obey excess entropy scaling as well as the above-mentioned Rosenfeld-Tarazona relation for the specific heat's temperature dependence~\cite{Rosenfeld1998}.  The implication is that the isomorph theory applies also to systems that are not conventional liquids, such as colloidal suspensions and dusty plasmas. In fact, the findings presented below confirm a recently derived theory of quasiuniversality according to which the Yukawa system is in the so-called \emph{EXP} quasiuniversality class of simple, monatomic, pair-potential  systems, a class that includes also, e.g., the inverse-power-law (IPL) and Lennard-Jones systems~\cite{Bacher2014a}.

In the following we first give a short overview of the isomorph theory and the computer simulation procedures used in the paper (Secs.~\ref{sec:theory} and~\ref{sec:simulations}). We proceed to show how to construct isomorphs using two different methods in Sec.~\ref{sec:isomorphs}. Sec.~\ref{sec:results} then shows that the isomorphs generated by these methods are very close to one another and confirms the predicted isomorph invariance of structure and dynamics. Finally, Sec.~\ref{sec:discussion} gives a brief discussion.

\section{Isomorphs in Roskilde-simple systems: A brief review\label{sec:theory}}

The isomorph theory uses so-called reduced units in which quantities are made dimensionless via units based on the thermodynamic quantities density and temperature, {\it not} the length and energy of the microscopic potential that are often used in reporting simulation results. Thus one uses the unit $l_0\equiv\rho^{-1/3}$ for length, $e_0\equiv k_BT$ for energy, and $t_0\equiv\rho^{-1/3}/\sqrt{k_BT/m}$ for time with $m$ being the particle mass (for Brownian dynamics a different time unit is used) \cite{paper4}. For instance, the collective position of the system's particles, $\vect{R} \equiv (\vect{r}_1, \ldots, \vect{r}_N)$, is expressed in reduced units as

\begin{equation}\label{eq:reduced_conf} 
  \tilde{\vect{R}} \equiv \rho^{1/3} \vect{R}\,,
\end{equation}
where the tilde here and henceforth denotes a reduced quantity. 

The isomorph theory is conveniently summarized in the expression~\cite{Dyre2014}
\begin{equation}\label{eq:hidden_scale_invariance}
U(\vect{R}) \cong h(\rho) \tilde{\Phi}(\tilde{\vect{R}}) + g(\rho)\,.
\end{equation}
Here $\tilde{\Phi}$ is a dimensionless, state-point independent function of the reduced configurations $\tilde{\vect{R}}$; the term $\tilde{\Phi}(\tilde{\vect{R}})$ controls the structure and dynamics in reduced units, but notably it contains no length or energy scales. The functions $h(\rho)$ and $g(\rho)$ both have units of energy. The physics of Eq. (\ref{eq:hidden_scale_invariance}) is that upon changing the density of a system, the potential-energy surface to a good approximation merely undergoes a linear, affine rescaling. We proceed to show how \eq{eq:hidden_scale_invariance} is used to derive the most important properties of R systems, their strong virial energy correlations and the existence of isomorphs. Before doing so we note that the isomorph theory was recently generalized by defining a Roskilde-simple system by the property that the order of potential energies is maintained for uniform scaling of configurations, i.e., by the condition $U(\vect{R}_a)<U(\vect{R}_b)\Rightarrow U(\lambda\vect{R}_a)<U(\lambda\vect{R}_b)$~\cite{Schroder2014}. This leads to slightly modified predictions, but overall the new isomorph theory is close to the original, which is the one used below.

The microscopic virial of Eq. (\ref{eq5}) is defined by~\cite{AllenTildesley}
\begin{equation}\label{eq:Wdef}
W(\vect{R})\equiv -\frac{1}{3}\,\vect{R}\cdot\nabla U(\vect{R})\,.
\end{equation}
It is easy to show that $W(\vect{R})$ characterizes the change of potential energy for a uniform scaling of space (i.e., leaving $\tilde{\vect{R}}$ intact) as follows \cite{Dyre2013b}

\begin{equation}
  W(\vect{R}) = \left(\frac{\partial U(\vect{R})}{\partial\ln\rho} \right)_{\tilde{\vect{R}}}\,.
\end{equation}
Combining this with \eq{eq:hidden_scale_invariance} and eliminating $\tilde{\Phi}(\tilde{\vect{R}})$ one finds that

\begin{equation}\label{eq:gamma3}
  W(\vect{R}) \cong \gamma(\rho) U(\vect{R}) + \phi(\rho)\,.
\end{equation}
Here 

\begin{equation}\label{eq:gamma_hrho}
  \gamma(\rho) \equiv \dfrac {\ud \ln h(\rho)} {\ud\ln\rho}
\end{equation}
and $\phi(\rho) = \ud g(\rho)/\ud\ln\rho - g(\rho)\gamma(\rho)$. From \eq{eq:gamma3} it follows that at constant density the fluctuations in $W$ and $U$ are correlated with linear-regression slope $\gamma$ given by

\begin{equation}\label{eq:gamma-fluctuations}
  \gamma \equiv \frac {\left<\Delta W \Delta U\right>} {\left<(\Delta U)^2\right>}\,.
\end{equation}
As shown in \fig{fig:UWscatter}(a) the Yukawa system indeed has strong $WU$ correlations (with slope $\gamma=2.14$ at the state point studied here). 

We proceed to show that \eq{eq:hidden_scale_invariance} implies the existence of isomorphs. First these curves in the thermodynamic phase diagram need to be defined. Consider two densities, $\rho_1$ and $\rho_2$, and two configuration $\vect{R}_1$ and $\vect{R}_2$ at these densities, respectively, with the same reduced coordinates, i.e., $\rho_1^{1/3}\vect{R}_1 = \rho_2^{1/3}\vect{R}_2 \equiv \tilde{\vect{R}}$. Applying \eq{eq:hidden_scale_invariance} to $U(\vect{R}_1)$ and $U(\vect{R}_2)$ and eliminating the common factor $\tilde{\Phi}(\tilde{\vect{R}})$ we find $\left[U(\vect{R}_1) -g(\rho_1)\right]/h(\rho_1) = \left[U(\vect{R}_2) -g(\rho_2)\right]/h(\rho_2)$. Define now two temperatures by $k_BT_1\equiv Kh(\rho_1)$ and $k_BT_2\equiv Kh(\rho_2)$ with the same proportionality constant $K$. This gives

\begin{equation}\label{eq:isomorphDef_log}
  \dfrac{U(\vect{R}_1)}{k_BT_1} \cong \dfrac{U(\vect{R}_2)}{k_BT_2} + B_{12}\,,
\end{equation}
where the constant $B_{12}$ depends only the two state points, not on the configurations. This can be rewritten as

\begin{equation}\label{eq:isomorphDef}
  \exp\left(-\dfrac{U(\vect{R}_1)}{k_B T_1}\right) \cong C_{12} \exp\left(-\dfrac{U(\vect{R}_2)}{k_B T_2}\right)\,. 
\end{equation}
This equation implies identical canonical probabilities of configurations with the same reduced coordinates. When this is obeyed to a good approximation for most of the physically relevant configurations, the two state points $(\rho_1,T_1)$ and $(\rho_2,T_2)$ are by definition {\it isomorphic} to one another~\cite{paper4}. This defines a mathematical equivalence relation in the thermodynamic phase diagram, and an isomorph is then defined as a curve along which all pairs of state points are isomorphic.

From the isomorph definition \eq{eq:isomorphDef} it follows that many properties of the system are invariant between isomorphic state points~\cite{paper4}. Isomorph invariants include thermodynamic quantities such as the excess entropy and the isochoric specific heat, as well as the reduced-unit dynamics and structure. Not all quantities are invariant even in reduced units, for instance the free energy and its volume derivatives like pressure or compressibility. 

Because the temperature is proportional to $h(\rho)$ along an isomorph with the same proportionality constant, isomorphs are described by

\begin{equation}\label{eq:T_hrho_isomorph}
  \frac{h(\rho)}{T}={\rm Const.}
\end{equation}
Since moreover the excess entropy per particle $s_{ex}$ is constant on an isomorph (because the Boltzmann probabilities of scaled configurations are), the temperature of an R system, i.e., one with isomorphs, separates as follows~\cite{Ingebrigtsen2012}

\begin{equation}\label{eq:factorization}
  k_BT = f(s_{ex})h(\rho)\,.
\end{equation}
It can be shown that this separation property is mathematically equivalent to the thermal average of \eq{eq:gamma3}, which is simply the configurational Gr{\"u}neisen equation of state $W\cong\gamma(\rho) U+\phi(\rho)$~\cite{Ingebrigtsen2012}.

The isomorph theory is approximate for all realistic systems. In fact, Eqs.(\ref{eq:hidden_scale_invariance}) and (\ref{eq:isomorphDef}) are only exact for systems that have an Euler-homogeneous potential-energy function. The most important example of such systems are those with an inverse-power-law (IPL) pair potential ($\upsilon(r) \propto r^{-n}$). IPL systems are easily shown to be characterized by $h(\rho)\propto \rho^\gamma$, and it follows from \eq{eq:Wdef} that $\gamma$ is related to the exponent $n$ of the IPL pair potential by

\begin{equation}\label{eq:gamma_ipl_exponent}
  \gamma = \frac{n}{3}\,.
\end{equation}
The applicability of the isomorph theory for simple atomic systems may be understood physically from the fact that, in the region of the first peak of the radial distribution function, the interatomic potential is well fitted by an IPL term plus a linear term, the so-called extended IPL (eIPL) pair potential~\cite{paper2}. This provides an intuitive explanation why many non-IPL liquids obey the isomorph theory. Notably, the linear term contributes little to the fluctuations in the energy or the virial. The reason for this is that when a particle moves, interparticle distance decrease on one side while increasing on the other side. The sum of the interparticle distances stays approximately the same. ~\cite{paper2}

\section{Simulation procedure\label{sec:simulations}}

Our simulations used the ``Roskilde University Molecular Dynamics'' (RUMD) code, which is optimized for GPU computing~\cite{RUMD}. All simulations were performed in the $NVT$ ensemble with a standard Nos{\'e}-Hoover thermostat. A state point is characterized by two parameters: the density $\rho$, reported below in the unit system defined by the screening length $\lambda$ of \eq{eq:y_debyuehuckel}, and the temperature $T$ reported in the unit system defined by the unit $Q^2/(\lambda k_B)$. Equivalently, a state point may be characterized by the two dimensionless numbers the screening parameter $\kappa$ of \eq{scr} and the coupling parameter $\Gamma$ of \eq{coup}. For maximum clarity the simulation results are presented both in terms of $\rho$ and $T$, and in terms of $\kappa$ and $\Gamma$. In practice the simulations were performed directly in reduced units, i.e., density and temperature were set equal to unity, changing instead the length and energy parameters of the potential in order to investigate different state points in the thermodynamic phase diagram. 

In reduced units the integration time step $\Delta \tilde{t}$ was decreased at low screening lengths ($\kappa\gg 1$) because the potential here becomes very steep. We used a shifted-force cutoff~\cite{Toxvaerd2011} for the potential, with a (reduced) cutoff $r_{cut}$ that varied with the state point because longer cutoffs are needed at low $\kappa$ values. Wherever necessary we increased the number of particles $N$ to allow for larger cutoffs. It should be noted that this cutoff method does not yield accurate values of the potential energy, especially at high densities close to the OCP limit. The accurate calculation of the potential energy of the OCP hase been discussed elsewhere~\cite{Hansen1973, Ng1974, Slattery1980}, and is beyond the scope of this investigation. Table~\ref{tab:y_simdetails} summarizes the simulation parameters used.

\begin{table}
  \centering
  \setlength{\tabcolsep}{6pt}
  \begin{tabular}{l|crr|l}
     & \multicolumn{1}{c}{$N$} &  \multicolumn{1}{c}{$\tilde{r}_{cut}$} & \multicolumn{1}{c|}{$\Delta\tilde{t}$ } & \\
    \hline
    $  \qquad\quad\;\! \rho < 0.01$ & 2048 & 4.3  & 0.0010 & $\;\;\:  5 < \kappa$       \\
    $0.01    \leqslant \rho < 0.5$  & 2048 & 5.2  & 0.0025 & $      1.5 < \kappa < 5$   \\
    $0.5\;\: \leqslant \rho$        & 8192 & 10.0 & 0.0025 & $\qquad\:\:\,\kappa < 1.5$ \\
    \hline
  \end{tabular}
  \caption{The parameters used in the simulations. Depending on the density $\rho$ the number of particles in the system $N$ was changed to allow for a larger reduced cutoff radius $\tilde{r}_{cut}$. The reduced time step $\Delta\tilde{t}$ was decreased at the lowest densities.}
  \label{tab:y_simdetails}
\end{table}

\section{Two methods for identifying the isomorphs\label{sec:isomorphs}}

This section details two methods for mapping out an isomorph in the thermodynamic phase diagram. The first one is numerical; it is accurate for small density variations, but not well suited for studies involving large density variation. Here a recently proposed approximate analytical method is more handy. This method estimates the isomorph from the pair potential by using \eq{eq:T_hrho_isomorph} in conjunction with an analytical expression for $h(\rho)$. This has been shown to work well for the Lennard-Jones system and two systems with pair potentials that are sums of two, respectively three IPL terms~\cite{Bohling2014}.

\subsection{The direct isomorph check: A numerical method\label{ssec:dic}}

Equation (\ref{eq:isomorphDef_log}) can be rewritten as
\begin{equation}\label{eq:dic}
  U(\vect{R}_2) \cong \dfrac{T_2}{T_1}U(\vect{R}_1) + D_{12}\,.
\end{equation}
This shows that the potential energies of configurations at density $\rho_1$ and the same ones scaled to density $\rho_2$ are predicted to be linearly related. If the configurations are taken from an equilibrium simulation at state point $(\rho_1,T_1)$, the slope is the ratio $T_2/T_1$, where $T_2$ is the temperature for which state point $(\rho_2,T_2)$ is isomorphic to state point $(\rho_1,T_1)$. This can be used to find state points that are isomorphic to each other. This procedure is termed the direct isomorph check because it is based directly on the isomorph definition \eq{eq:isomorphDef}~\cite{paper4}. 

\begin{figure}
  \centering
  \includegraphics[width=\figwidth]{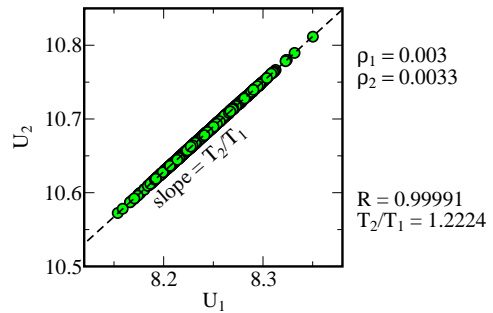}
  \caption{The Yukawa system simulated at $\rho_1 = 3\times 10^{-3}$ ($\kappa=6.93$) and $T_1=5.361\times 10^{-5}$ ($\Gamma=2690$). The potential energies $U_1\equiv U(\vect{R_1})$ of configurations were plotted versus the potential energies $U_2\equiv U(\vect{R_2})$ of the same configurations scaled to a 10\% higher density denoted $\rho_2$ (green symbols). The two energies are highly correlated, indicating that the Yukawa system obeys the isomorph definition (\eq{eq:isomorphDef_log}). By simple linear regression (dashed line) we find the slope $T_2/T_1$ (see \eq{eq:dic}), which means that for the state point at $\rho_1$ to be isomorphic to the one at $\rho_2$, the temperature of the latter should be $T_2=1.2224\times T_1=6.553\times 10^{-5}$ ($\Gamma=2272$).}
  \label{fig:UUscatter}
\end{figure}

The direct-isomorph-check procedure is illustrated in \fig{fig:UUscatter}. Here a number of configurations $\vect{R}_1$ with potential energy denoted by $U_1$ were sampled from an equilibrium simulation of the Yukawa fluid at density $\rho_1$. Each configuration was then scaled to the higher density $\rho_2=1.1\times\rho_1$, at which the new potential energy $U_2$ was calculated. The figure shows that the potential energies $U_1$ and $U_2$ are highly correlated. The standard linear-regression slope is $1.2224$. According to \eq{eq:dic} this number is the ratio $T_2/T_1$, which allows for determining $T_2$ such that the state point $(\rho_2,T_2)$ is isomorphic to $(\rho_1,T_1)$. We see that along the isomorph through $(\rho_1,T_1)$ a density increase of 10\% implies a 22.2\% increase in temperature.

The structure and dynamics of the two isomorphic state points are shown to be the same in \fig{fig:smallstep}, indicating that the direct isomorph check works well for density changes of 10\%. However, it should be noted that the change in density cannot be huge, because the isomorph theory is only approximate (except for IPL systems), which implies that direct-isomorph-check plots give relatively poor correlations for large density jumps. To avoid this problem, we used in this work always a density change of merely 10\% for direct isomorph checks, corresponding to changing $\kappa$ by less than 4\%. In the simulations presented in section~\ref{sec:results} we used the direct isomorph check to create an isomorph by doing a simulation at the initial state point $(\rho,T) = (10^{-3},3\times 10^{-6})$, scaling configurations to a new density, and finding the temperature of the new isomorphic state point. This was repeated at the new state point to obtain a third one, etc. In this way we obtained altogether a set of 32 prospective isomorphic state points in the range $3\times 10^{-4}\leqslant\rho\leqslant 3.6\times 10^{-3}$ ($6.5\leqslant\kappa\leqslant 14.9$). These state points are referred to here as ``prospective'' because they will be compared to other sets of prospective isomorphic state points generated in different ways.

\subsection{Predicting the isomorph analytically from the pair potential}

In this section we aim to construct an isomorph by obtaining an expression for the function $h(\rho)$ of \eq{eq:hidden_scale_invariance}. For pair potentials that are a sum of IPLs, $h(\rho)$ can be determined from a single simulation at a reference state point~\cite{Ingebrigtsen2012, Bohling2012}. For other potentials such as the Yukawa this is not possible. Nevertheless, B{\o}hling {\it et al.}~\cite{Bohling2014} have recently shown that $h(\rho)$ can be estimated from the potential. We briefly review these findings before applying this method to the Yukawa potential.

\begin{figure}
  \centering
  \includegraphics[width = \figwidth]{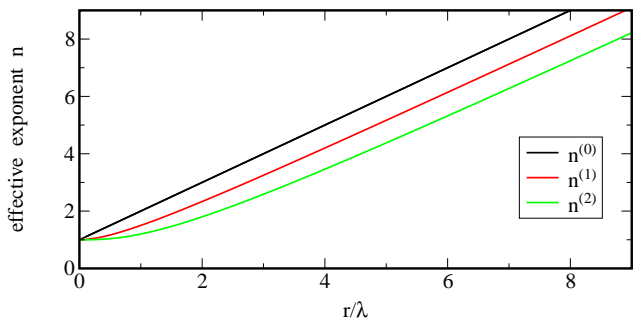}
  \caption{Estimates of the effective IPL exponent $n^{(p)}(r)$ for the Yukawa potential using \eq{eq:n_general}.}
  \label{fig:y_n2-r}
\end{figure}

As mentioned earlier, for an IPL pair potential $\propto r^{-n}$ it is known that $\gamma=n/3$ (\eq{eq:gamma_ipl_exponent}). For other potentials an effective $r$-dependent IPL exponent can be estimated using ratios of derivatives of the potential~\cite{paper2} as follows:
\begin{equation}\label{eq:n_general}
  n^{(p)}(r) \equiv -r\frac{\upsilon^{(p+1)}(r)}{\upsilon^{(p)}(r)} -p\,,
\end{equation}
where $\upsilon^{(p)}$ denotes the $p$th derivative of the potential. For an IPL pair potential, $n^{(p)}(r)$ is constant and gives the correct exponent for all $p$. For other potentials, the effective exponent depends on both $p$ and $r$, meaning that the ``softness'' of the particles depends on the separation between the particles. We show this in \fig{fig:y_n2-r} for the Yukawa potential. It is known that in the OCP limit the potential reduces to a Coulomb interaction, for which $n=1$. This corresponds to small interparticle distances and we see indeed that in the limit of $r\rightarrow 0$ one finds $n^{(p)}(r)\rightarrow 1$ for every $p$. 

Recall that potentials that obey the isomorph theory have been found to be well fitted by an extended IPL potential, i.e., an IPL potential plus a linear term~\cite{paper2}.
An obvious choice for $p$ is thus 2, since this would ignore the linear contribution to the extended IPL potential:
\begin{equation}\label{eq:n2}
  n^{(2)}(r) \equiv -r\frac{\upsilon'''(r)}{\upsilon''(r)} -2\,.
\end{equation}
The next question is at which distance to evaluate $n^{(2)}(r)$. Since the physics of R liquids has been shown to be governed by the interactions in the first coordination shell~\cite{Ingebrigtsen2012c}, the nearest neighbor distance is an obvious choice. Since any distance scales with density as $r\propto\rho^{-1/3}$, we can write the nearest-neighbor distance as $\Lambda\rho^{-1/3}$, where $\Lambda$ is a number close to unity. By combining \eq{eq:n2} and \eq{eq:gamma_ipl_exponent} one finds~\cite{Bohling2014}
\begin{equation}\label{eq:gamma_potential2}
  \gamma(\rho) = \left.\dfrac{n^{(2)}(r)}{3}\right|_{r=\Lambda\rho^{-1/3}}\,.
\end{equation}
 As noted by B{\o}hling {\it et al.}~\cite{Bohling2014}, $\Lambda$ is not expected to be the same for all state points. However, because structure in reduced units is predicted to be invariant along an isomorph, $\Lambda$ must also be an isomorph invariant. By comparing with simulation results of different pair potentials, they indeed find that $\gamma(\rho)$ does not predict the density-scaling exponent equally well for different isomorphs if $\Lambda$ is isomorph independent, although the prediction is qualitatively correct. Instead, an isomorph dependent (and thus excess entropy  dependent) nearest-neighbor distance $\Lambda(s_{ex})\rho^{-1/3}$ was able to estimate the isomorph more precisely, giving~\cite{Bohling2014}
\begin{equation}\label{eq:gamma_potential}
  \gamma(\rho,s_{ex}) = \left.\dfrac{n^{(2)}(r)}{3}\right|_{r=\Lambda(s_{ex})\rho^{-1/3}}\,.
\end{equation}

It is possible to rewrite \eq{eq:n2} as $n^{(2)}(r) \equiv \ud \ln[r^2\upsilon''(r)] / \ud \ln r$. This can be combined with \eq{eq:gamma_hrho} to find~\cite{Bohling2014}
\begin{equation}\label{eq:hrho_potential}
  h(\rho,s_{ex}) = \left.Ar^2\upsilon''(r)\right|_{r=\Lambda(s_{ex})\rho^{-1/3}}\,,
\end{equation}
where $A$ is an arbitrary constant. This can now be used to find an isomorph in the phase diagram using the facts that $h(\rho,s_{ex})/T={\rm Const.}$ (\eq{eq:T_hrho_isomorph}) and $s_{ex}$ is an isomorph invariant.

Applying this method to the Yukawa potential, we find using \eq{eq:hrho_potential} after straightforward calculations
\begin{equation}\label{eq:y_hrho}
  h(\rho,s_{ex}) = A\exp^{-\Lambda\rho^{-1/3}} \left[\Lambda\rho^{-1/3} + 2 + 2\Lambda^{-1}\rho^{1/3}\right]\,.
\end{equation}
If we express $h(\rho,s_{ex})$ in terms of the Yukawa parameters $\kappa$ and $\Gamma$, we find
\begin{equation}\label{eq:Gamma-k}
  \Gamma = \Gamma_0 \dfrac{2e^{\Lambda\kappa}}{(\Lambda\kappa)^2 + 2\Lambda\kappa + 2}\,
\end{equation}
where $\Gamma_0=\Lambda/(2A)$ is the value of $\Gamma$ in the OCP limit ($\kappa=0$). Vaulina {\it et al.}~\cite{Vaulina2000} found that a curve of this shape with $\Lambda=1$ and $\Gamma_0=106.6$ gives a good description of the melting line as found by Hamaguchi {\it et al.}~\cite{Hamaguchi1997}, and also of the melting line of dissipative Yukawa systems~\cite{Vaulina2002}. Their findings are fully consistent with the isomorph theory and now placed in a more general setting: The melting line is predicted to be an isomorph~\cite{paper4} since if an isomorph were to cross the melting line this would mean \eq{eq:isomorphDef} should be obeyed for a pair of liquid and solid state points. This cannot be the case, because the relative Boltzmann probabilities for liquid and solid configurations are clearly different at these two state points. Vaulina {\it et al.} used Lindemann's melting criterion which states that melting happens when the root mean square displacement is 10\% of the crystal's nearest neighbor distance~\cite{Gilvarry1956}. Since both structure and dynamics are invariant on the isomorph, they should also be so at the melting line. Lindemann's melting criterion is thus consistent with the isomorph theory~\cite{paper4}. It should also be noted that Yazdi {\it et al.}~\cite{Yazdi2014} recently showed that \eq{eq:Gamma-k} with $\Gamma_0=368$ and $\Lambda=1$ gives a good description of the ideal glass transition of mode-coupling theory; apparently this line is an isomorph which makes sense since isomorphs are lines of identical physics (in reduced units). A consequence of the melting line being an isomorph is also that the temperature $T$ on an isomorph scaled by the melting temperature $T_m$ is an isomorph invariant if isomorphs have the same shape (i.e., their shape is given by $h(\rho)$ instead of $h(\rho,s_{ex}))$. Thus so-called melting temperature scaling methods that have found that dynamic and/or thermodynamic properties of the Yukawa system~\cite{Rosenfeld2000, Rosenfeld2001, Khrapak2015a, Khrapak2015b} and the OCP~\cite{Rosenfeld2000, Rosenfeld2001, Khrapak2013} are functions only of $T/T_m$, are saying that these properties are isomorph invariants.

In addition to the direct isomorph check described in section~\ref{ssec:dic} we used \eq{eq:y_hrho} to obtain two sets of isomorphic state points. For this it is necessary to know the relevant reduced interparticle distance $\Lambda$. Because Vaulina {\it et al.}~\cite{Vaulina2000} found \eq{eq:Gamma-k} with $\Lambda=1$ to be a good description of the melting line and the melting line is an isomorph, we test if \eq{eq:y_hrho} with $\Lambda=1$ is the correct description of an isomorph. The prospective isomorph with $\Lambda=1$ that we have tested is characterized by $A=6.347\times 10^{-3}$ ($\Gamma_0 = 78.78$).

We also tried to calculate a more accurate value of $\Lambda$. Previously, B{\o}hling {\it et al.}~\cite{Bohling2014} used the most probable nearest neighbor distance for $\Lambda$, which they determined from the position of the first peak in $r^2g(r)$, with $g(r)$ being the radial distribution function. They found for the potentials they tested at different state points that $0.975\leqslant\Lambda\leqslant 1.065$. We below use the different method to determine $\Lambda$ from the $U,W$ fluctuations published earlier by Bailey {\it et al.}~\cite{Bailey2013}. First, we derive the expression for $\gamma(\rho,s_{ex}) = \partial \ln h(\rho,s_{ex})/\partial \ln \rho$ (\eq{eq:gamma_hrho}) for the Yukawa potential using \eq{eq:y_hrho}:
\begin{equation}\label{eq:y_gamma_rho_isomorph}
  \gamma(\rho,s_{ex}) = \dfrac {\Lambda^3} {3\Lambda^2\rho^{1/3} + 6\Lambda\rho^{2/3} + \rho} + \dfrac{1}{3}\,.
\end{equation}
We then did a simulation at an initial state point ($\rho=0.005$ and $T=0.00015$). At this state point $\gamma$ was found from the fluctuations to be 1.78 using \eq{eq:gamma-fluctuations}. From \eq{eq:y_gamma_rho_isomorph} it then follows that $\Lambda=1.03$ is the relevant reduced nearest-neighbor distance at that state point. The third prospective isomorph is thus identified by \eq{eq:y_hrho} with $\Lambda=1.03$ and $A=7.414\times 10^{-3}$ ($\Gamma_0=69.46$) to obtain a set of state points with the same densities as the points generated for the $\Lambda=1$ prospective isomorph.

\section{Comparing the three prospective isomorphs\label{sec:results}}
\begin{figure}
  \centering
  \includegraphics[width=\figwidth]{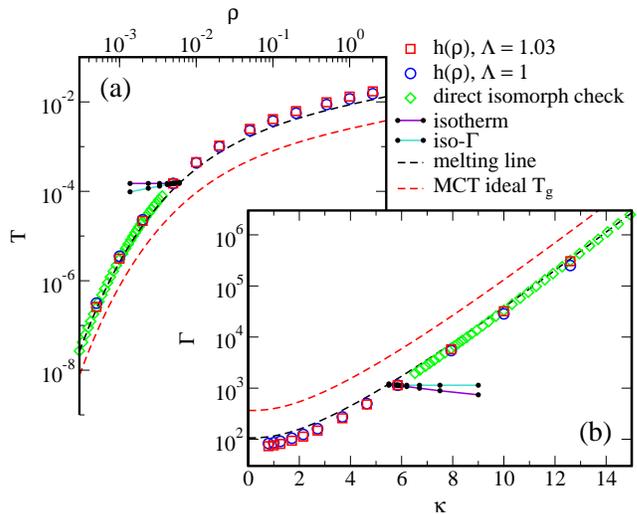}
  \caption{Prospective isomorphs in the (a) $\rho,T$ phase diagram and the (b) $\kappa,\Gamma$ phase diagram generated in the three different ways described in Sec. IV. Comparing the two estimates of $h(\rho)$ from the pair potential, there is a visible difference in the slope at low density (high $\kappa$). The isomorphs are approximately parallel to the expressions of the melting line~\cite{Vaulina2000} and the ideal MCT glass transition~\cite{Yazdi2014} (dashed lines), showing that these lines are both isomorphs. This is what one expects from the fact that the physics is invariant along the isomorphs. Sets of isothermal and iso-$\Gamma$ state points have also been included in the figure because these are used later in this section.}\label{fig:y_im_RhoT}
\end{figure}

This section compares the three sets of prospective isomorphic state points generated as described in the previous section. The three prospective isomorphs are shown in \fig{fig:y_im_RhoT} plotted both in the $\rho,T$ and the $\kappa,\Gamma$ plane. The isomorphs are parallel to the expression for the melting line~\cite{Vaulina2000} and the ideal glass transition line from mode coupling theory~\cite{Yazdi2014}. The three prospective isomorphs are slightly different, but overall close to one another. It is therefore not surprising that they have similar invariance properties. The remainder of this section focuses on illuminating the minor differences.

\begin{figure}
  \centering
  \includegraphics[width=\figwidth]{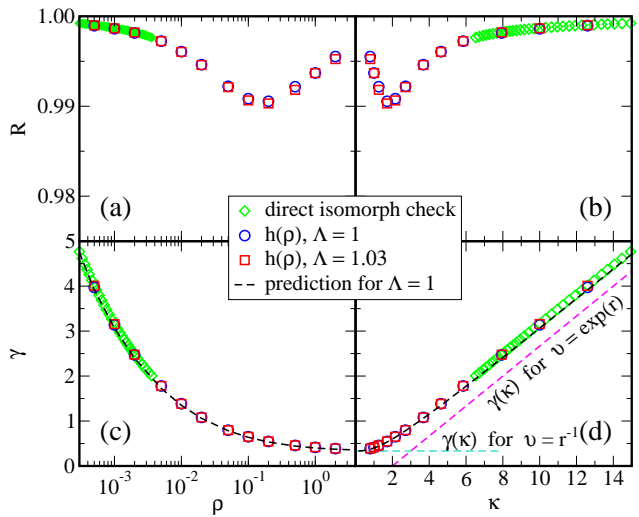}
  \caption{The virial potential-energy correlation coefficient $R$ (top), and the so-called density-scaling coefficient $\gamma$ (bottom) plotted versus density (left) and the screening parameter $\kappa$ (right). The Yukawa fluid is Roskilde-simple at all simulated state points. We find that $\gamma$ varies significantly, reflecting a dramatic change in steepness of the Yukawa potential with density.}\label{fig:y_im_Rgamma}
\end{figure}

Figures \ref{fig:y_im_Rgamma}(a) and (b) verify that the strong correlations between the virial and the potential-energy $NVT$ equilibrium fluctuations are present over the entire range of densities studied. Whereas models of conventional liquids are usually not strongly correlating at low densities and temperatures due to the large contribution of the attractive term in the potential, the Yukawa system is in fact strongly correlating with correlation coefficient $R>0.99$ (\eq{eq:R}) at all tested state points. The values $R>0.99$ are very high, especially when compared to those of other models that, as mentioned, are considered Roskilde liquids if they  obey $R>0.90$~\cite{paper1, paper2, Veldhorst2012, Ingebrigtsen2012b}. Recently, the ten-bead rigid-bond flexible Lennard-Jones chain was even demonstrated to have excellent isomorphs despite having $R\approx 0.86$~\cite{Veldhorst2014}. 

Figures \ref{fig:y_im_Rgamma}(c) and (d) show the density-scaling coefficient $\gamma$ calculated using \eq{eq:gamma-fluctuations}. 
In the OCP limit we find that $\gamma$ goes to $1/3$ as expected for an IPL with $n=1$. At lower densities the behavior of $\gamma$ is influenced by the exponential term, for which the estimated $\gamma(\rho)$ has also been shown for comparison. Due to the increase in the steepness of the potential, $\gamma$ increases from $1/3$ to 5 in the density range shown. The lowest correlation coefficient is found in the crossover region of densities marking the region where the effects of both the exponential and Coulomb terms are important.

\begin{figure}
  \centering
  \includegraphics[width=\figwidth]{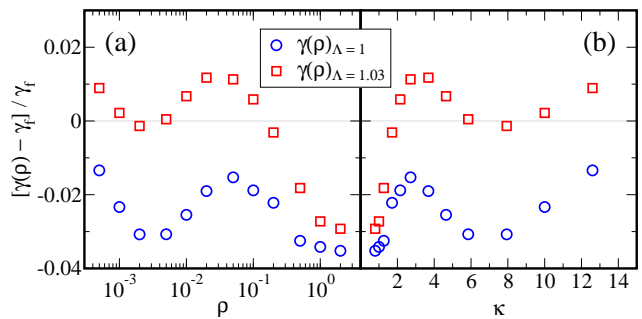}
  \caption{Relative difference between the estimated $\gamma(\rho)$ (\eq{eq:y_gamma_rho_isomorph}) and $\gamma$ calculated from the fluctuations via \eq{eq:gamma-fluctuations} (denoted in the figure by $\gamma_f$). The prediction $\gamma(\rho)_{\Lambda=1.03}$ is generally in best agreement with the values from the equilibrium fluctuations.}
  \label{fig:im23_gamma-diff}
\end{figure}

There is a slight difference between the values of $\gamma$ calculated from the fluctuations and those predicted for $\gamma(\rho,s_{ex})$ from \eq{eq:y_gamma_rho_isomorph} with $\Lambda=1$ (black dashed line). To investigate this further we plot in \fig{fig:im23_gamma-diff} the relative difference between the two estimated functions $\gamma(\rho,s_{ex})$ for $\Lambda=1$ and $\Lambda=1.03$ and the values of $\gamma$ calculated from the fluctuations. Both predictions are too low at high density (small $\kappa$). Unsurprisingly, the prediction with $\Lambda=1.03$, a number that was identified using the value of $\gamma$ from the energy and virial fluctuations at a reference state point with $\rho=0.005$ ($\kappa=5.85$), is more accurate at low densities (large $\kappa$).

\begin{figure}
  \centering
  \includegraphics[width=\figwidth]{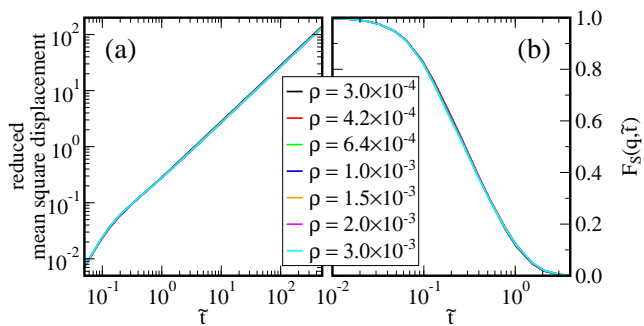}
  \caption{(a) The reduced mean-square displacement and (b) the self intermediate scattering function plotted as functions of the reduced time on the isomorph generated using the direct isomorph check. The dynamics collapse almost perfectly. The $q$ vector is kept constant in reduced units as $q=\rho^{1/3}\tilde{q}$ with $\tilde{q}=7$.}\label{fig:im8_dynamics}
\end{figure}

\begin{figure}
  \centering
  \includegraphics[width=\figwidth]{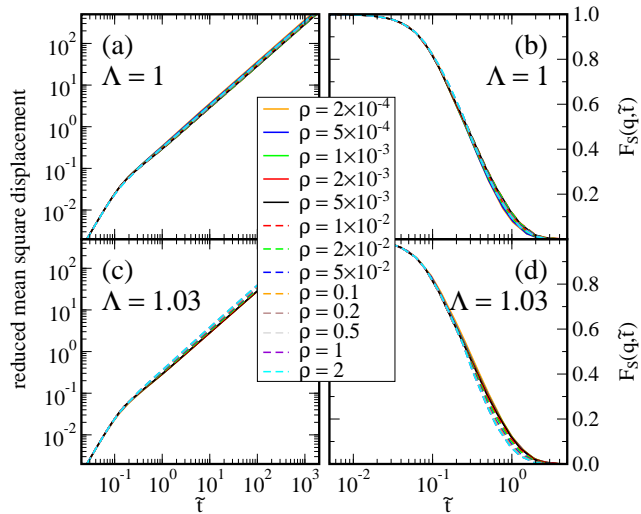}
  \caption{(a) and (c) The reduced mean-square displacements (left); (b) and (d) intermediate scattering functions (right) along the two isomorphs obtained from \eq{eq:y_hrho} for $\Lambda=1$ (top) and $\Lambda=1.03$ (bottom). Both sets of state points show almost invariant dynamics, although for $\Lambda=1.03$, the collapse is slightly worse than for $\Lambda=1$. The reduced scattering vector was again $\tilde{q}=7$.}\label{fig:im23_dynamics}
\end{figure}

We proceed to check to which degree the three sets of prospective isomorphic state points exhibit the invariance of dynamics and structure predicted for isomorphs.
We test the invariance of the dynamics in \fig{fig:im8_dynamics} for the isomorph generated by the direct-isomorph-check method and in \fig{fig:im23_dynamics} for the two isomorphs generated from the estimates of $h(\rho)$. The figures show the mean-square displacements and the intermediate scattering functions in reduced units. For all three sets of prospective isomorphic state points both measures of the dynamics collapse nicely onto a single curve. 

\begin{figure}
  \centering
  \includegraphics[width=\figwidth]{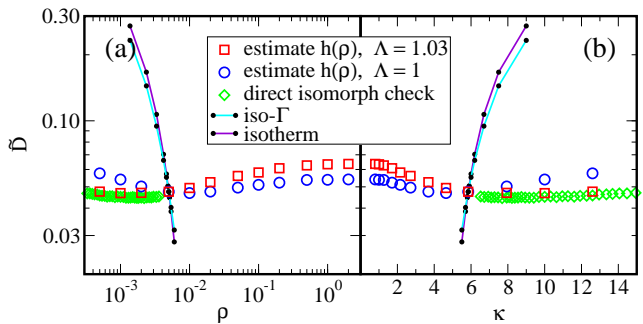}
  \caption{The reduced diffusion coefficient along the three prospective isomorphs, calculated from linear regression of the part of the mean-square displacement above unity. The diffusion coefficients of the state points obtained with the direct isomorph check are constant in the tested range of densities. The two estimates from \eq{eq:Gamma-k} have been tested in much larger range, and show a larger deviation from isomorph invariance, especially the estimate with $\Lambda=1.03$ at high density. Overall, the dynamics on either of the three prospective isomorphs are much more invariant, however, than on the curves of constant temperature or constant $\Gamma$.}\label{fig:diffusion}
\end{figure}

There are only minor differences in how invariant the dynamics are on the three prospective  isomorphs. To amplify these differences we calculated the reduced diffusion coefficient from the mean-square displacement at all investigated state points. The results are shown in \fig{fig:diffusion}, and compared with the diffusion coefficient along an isotherm and a curve of constant $\Gamma$. In view of the large density range simulated there is  little variation in the diffusion coefficient along the isomorph. The isomorph obtained with the direct-isomorph-check method only covers part of the phase diagram (because of the time-consuming nature of obtaining these state points). The reduced diffusion coefficients are virtually constant over the range simulated by this method. The results for the isomorph estimated from the pair potential with $\Lambda=1.03$ collapse with those of the isomorph from the direct isomorph check. Interestingly, the range where the diffusion coefficients are almost invariant coincides with the range where agreement between the estimated and the fluctuation $\gamma$ is best. We also note that although $\Lambda=1$ leads to a worse prediction for $\gamma$, it gives a more invariant diffusion coefficient when the whole phase diagram is considered. We attribute this to some cancellation of errors, rather than reflecting that $\Lambda=1$ gives a more precise value of the relevant interparticle distance. Note also that the invariance of the dynamics, and thus the isomorphs seems to continue all the way to the OCP limit. As mentioned earlier, also the melting line is an isomorph, so our results indicate that also the melting line continues to the OCP limit, indicating the existence of a phase transition in the OCP~\cite{Stishov1998}, at least from a dynamical point of view.

\begin{figure}
  \centering
  \includegraphics[width=\figwidth]{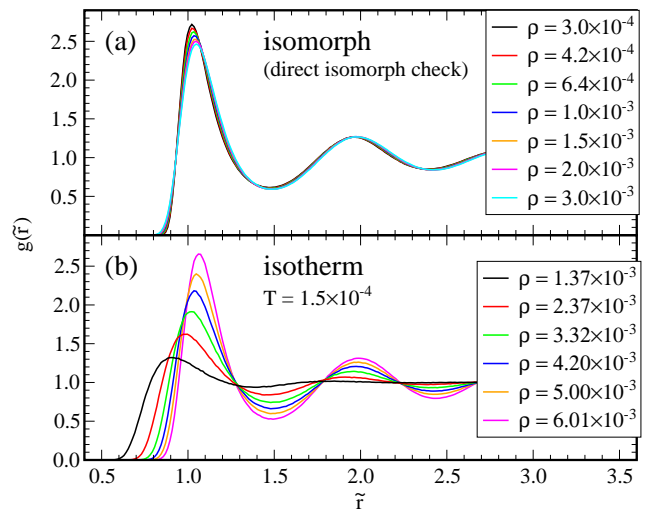}
  \caption{The reduced radial distribution function $g(\tilde{r})$ at isomorphic state points obtained with the direct isomorph check (a). Although the change in density is an entire decade, the reduced-unit structure does not change significantly along the isomorph. The invariance is especially striking when compared to the reduced radial distribution functions of isothermal state points (b), especially considering the smaller density range for the isotherm.}\label{fig:y_im8_structure}
\end{figure}

\begin{figure}
  \centering
  \includegraphics[width=\figwidth]{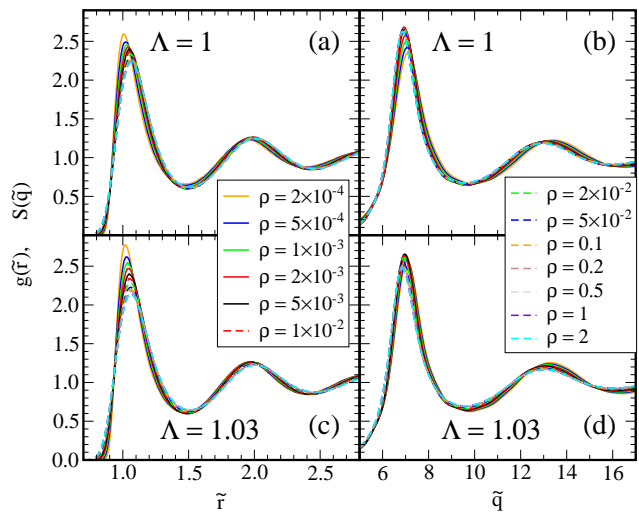}
  \caption{The radial distribution functions (left) and static structure factors (right) for state points obtained from \eq{eq:y_hrho} for two values of $\Lambda$. The structure is not completely invariant on the isomorph; both the height and the position of the first peak vary somewhat.}\label{fig:y_im23_rdf}
\end{figure}

We proceed now to testing the isomorph invariance of structure as quantified via the radial distribution function $g(\tilde{r})$ and the static structure factor $S(\tilde{q})$ (both in reduced units). Results for the prospective isomorphic state points obtained using the direct isomorph check are shown in \fig{fig:y_im8_structure}; results for state points of the two estimates of $h(\rho)$ with $\Lambda=1$ and $\Lambda=1.03$ are plotted in \fig{fig:y_im23_rdf}. Overall, the structure is invariant on the three prospective isomorphs when compared to the change in $g(\tilde{r})$ along the isotherm. This is especially impressive considering the density on the isotherm changes by a factor of 4.4, while on the isomorphs it changes by a factor of 10 (\fig{fig:y_im8_structure}(a)) and $10^4$ (\fig{fig:y_im8_structure}).

Whereas the 10\% density change applied in \fig{fig:smallstep} resulted in invariance of the structure, the large density changes applied here (a factor of 10 and $10^4$ respectively) \emph{do} lead to systematic changes in the structure along the proposed isomorph, especially in the region of the first peak of $g(\tilde{r})$. This is not unexpected, because the increase in the steepness of the potential with decreasing density (see \fig{fig:y_im_Rgamma}) makes close encounters between particles less probable. This results in a steeper initial slope of $g(\tilde{r})$ and thus a higher first peak if the total number of nearest neighbors is unchanged ~\cite{Bohling2014}. Taking into account that the direct-isomorph-check isomorph ``only'' involves densities variations covering a single decade, we note that the deviation from isomorphic invariance is smallest for the state points obtained using $\Lambda=1$.

Not only the height of the peak, but also its position changes somewhat along the prospective isomorphs. This is an effect of the large change in the effective steepness of the potential (i.e., relative to $k_BT$). The Yukawa potential has for instance both a BCC and an FCC solid phase with the triple point at $\kappa=6.90$~\cite{Hamaguchi1997}. These two crystal structures have different reduced nearest-neighbor distances, and it is not surprising that some of these structural differences continue to exist in the liquid state. The change of the nearest-neighbor distance with density,  however, indicates that a single value for $\Lambda$ is not enough for accurately  tracing out the isomorph.

\vspace{1cm}

\section{Discussion\label{sec:discussion}}

We have shown that the Yukawa system is Roskilde simple. This was shown from the strong correlations between equilibrium virial and potential-energy fluctuations and by the fact that the Yukawa system has isomorphs in its phase diagram. An isomorph is a curve of constant excess entropy, but the curve can also be estimated directly from the pair potential. We have verified that the dynamics are invariant on the isomorphs as predicted by the theory. The structure of the fluid as characterized by the radial distribution function was found to be invariant to a lesser degree, in part because the position of the first peak shifts slightly. Our estimate of the isomorph shape from the pair potential uses the nearest-neighbor distance. Consequently, \eq{eq:y_gamma_rho_isomorph} with a constant $\Lambda$ can \emph{not} give an exact description of the density dependence of $\gamma$ along an isomorph. In view of the invariant dynamics shown in \fig{fig:im23_dynamics}(a) and (b), however, \eq{eq:y_hrho} with $\Lambda=1$ must be said to give a rather good description of the isomorph shape. 

In summary, we have shown that the isomorph theory provides a simple and general framework for understanding previous findings for the Yukawa system. Thus our results explain the previously identified expression of the melting line of Yukawa systems~\cite{Vaulina2000, Vaulina2002}, which here follows from the fact that the melting line is an isomorph~\cite{paper4}. Likewise, the isomorph theory~\cite{Robbins1988, Rosenfeld2000, Rosenfeld2001} explains the observation that the transport coefficients of the Yukawa fluid collapse when plotted versus temperature scaled by the melting temperature (compare \eq{eq:T_hrho_isomorph}), as well as the recently proposed expression for the ideal glass transition line of mode coupling theory~\cite{Yazdi2014}.Finally, we point out that the Yukawa system belongs to the ``exponential'' class of quasiuniversal systems~\cite{Bacher2014a}.

\acknowledgments{
The centre for viscous liquid dynamics ``Glass and Time'' is sponsored by the Danish National Research Foundation via Grant No. DNRF61.
}


\end{document}